\documentclass{emulateapj}

\usepackage{subfigure}
\usepackage{url}
\usepackage{hyperref}
\usepackage{datetime}
\usepackage{longtable}
\usepackage{mathrsfs}
\usepackage{amsbsy}

\newcommand{\noprint}[1]{}

\usepackage{color}

\newcommand{\rprs}{{$R_p/R_{\star}$}}

\newcommand{\eg}{{\it e.g.}}
\newcommand{\ie}{{\it i.e.}}
\newcommand{\itk}{{\it Kepler}}

\newcommand{\teff}{$T_{\rm eff}$}

\newcommand{\rearth}{{$R_\oplus$}}
\newcommand{\rsun}{{$R_\odot$}}
\newcommand{\msun}{{$M_\odot$}}

\newcommand{\mh}{{[M/H]}~}

\begin{document}
\title{Characterizing the Cool KOIs VIII. Parameters of the Planets Orbiting Kepler's Coolest Dwarfs}

\author{Jonathan J. Swift\altaffilmark{1,*}, Benjamin T. Montet\altaffilmark{1,2}, Andrew Vanderburg\altaffilmark{2}, Timothy Morton\altaffilmark{3}, Philip S. Muirhead\altaffilmark{4}, John Asher Johnson\altaffilmark{2}}

\altaffiltext{1}{California Institute of Technology, 1200 E. California Blvd., Pasadena, CA  91125 USA}
\altaffiltext{*}{Current address: The Thacher School, 5025 Thacher Rd., Ojai, CA 93023}
\altaffiltext{2}{Harvard-Smithsonian Center for Astrophysics, Cambridge, MA 02138 USA}
\altaffiltext{3}{Department of Astrophysical Sciences, Princeton University, 4 Ivy Lane, Peyton Hall, Princeton, NJ 08544, USA}
\altaffiltext{4}{Department of Astronomy, Boston University, 725 Commonwealth Ave., Boston, MA 02215 USA}
\date{\today, \currenttime}

\begin{abstract}
The coolest dwarf stars targeted by the \itk\ Mission constitute a relatively small but scientifically valuable subset of the \itk\ target stars, and provide a high-fidelity and nearby sample of transiting planetary systems. Using archival \itk\ data spanning the entire primary mission we perform a uniform analysis to extract, confirm and characterize the transit signals discovered by the \itk\ pipeline toward M--type dwarf stars. We recover all but two of the signals reported in a recent listing from the Exoplanet Archive resulting in 165 planet candidates associated with a sample of 106 low-mass stars. We fitted the observed light curves to transit models using Markov Chain Monte Carlo and we have made the posterior samples publicly available to facilitate further studies. We fitted empirical transit times to individual transit signals with significantly non-linear ephemerides for accurate recovery of transit parameters and measuring precise transit timing variations. We also provide the  physical parameters for the stellar sample, including new measurements of stellar rotation, allowing the conversion of transit parameters into planet radii and orbital parameters. 
\end{abstract}


\maketitle

\section{Introduction}
\label{sec:intro}
NASA's \itk\ Space Mission was designed to monitor more than 150,000 stars within a single 115 square degree patch of sky in search of periodic diminutions of light caused by transiting exoplanets \citep{Borucki2010,Koch2010,Jenkins2010}. \itk's great success in discovering transiting exoplanets \citep{Borucki2011a,Borucki2011b,Batalha2013,Burke2014} has revealed that planets are at least as numerous as stars in the Galaxy \citep{Fressin2013,Petigura2013b,Swift2013,Dressing2013,Morton2014a}. Beyond the sheer numbers of planets, \itk\ has also provided important insights into the characteristics of the transiting planet population. The multi-transit systems reveal highly coplanar multi-planet systems \citep{Lissauer2011b,Tremaine2012,Fang2012,Fabrycky2014, Ballard2014}, many of which are in compact configurations \citep[{\eg}][]{Lissauer2011a,Muirhead2012a,Swift2013}. The period ratios of adjacent transiting planets show an excess just outside of mean motion resonance \citep{Lissauer2011b,Fabrycky2014} that may reflect the mechanisms by which these systems formed \citep{Rein2012,Goldreich2014}, or else may indicate subsequent evolution of these systems \citep{Lithwick2012a,Batygin2013}. The typical surface density profile of the protoplanetary disks from which these planets formed can be estimated using the \itk\ sample, and implies that either protoplanetary disks contain a large amount of material within $\sim 0.1$\,AU of the host star \citep{Chiang2013,Hansen2012} or that the planets migrated from their birth places further out in the disk \citep{Swift2013,Schlichting2014}. Another clue regarding the formation mechanisms behind the \itk\ planet sample is the radius function---the frequency of planets as a function of their size---that shows unambiguously that there are many more planets with radii less than that of Neptune than there are larger ones \citep{Howard2012,Fressin2013,Petigura2013a,Morton2014a,Foreman-Mackey2014}.

Although the vast majority of \itk\ target stars are Sun-like ($0.8\,M_\odot \lesssim M_\star \lesssim 1.2\, M_\odot$), several thousand M dwarfs have been monitored by \itk\ over the course of the primary mission. The initial photometric characterization of the M dwarfs in the \itk\ field was known to be inaccurate because the the \itk\ Input Catalog was optimized for Sun-like stars  \citep[KIC;][]{Brown2011}. However, there have been several efforts to revise the stellar parameters for this sample \citep[{\eg}][]{Muirhead2012a,Mann2012,Mann2013c,Dressing2013,Muirhead2014,Newton2014}. Since the physical parameters of a transiting planet are dependent on the stellar parameters, many exciting results have come from a careful examination of this stellar sample \citep[{\eg}][]{Johnson2011a, Johnson2012,Muirhead2012b,Muirhead2013}. The depth of a transit signal is proportional to the square of the relative planet radius, $\delta \propto (R_p/R_\star)^2$ allowing the detection of smaller planets around these smaller stars. This higher sensitivity to smaller planets allows the planet population around \itk's M dwarfs to be well-sampled down to $\lesssim 1$\,\rearth, where planets are most prevalent \citep{Morton2014a}. Further, the notional ``habitable zone", in which planets have equilibrium temperatures comparable to that of the Earth, is much closer to these cool, faint stars. This increases the transit probability and number of transits per observing time baseline, thereby allowing the first detection and measurement of the occurrence of Earth--sized planets in the habitable zones of stars \citep{Dressing2013,Quintana2014}

As a supplement to our recent efforts to characterize the lowest mass stars in the \itk\ field \citep{Muirhead2012b,Muirhead2014}, we here focus on the transit signals in the list of M dwarf \itk\ Objects of Interest (KOIs). Following is a uniform treatment of the sample with which we derive a statistically useful body of information regarding the properties of the planets orbiting \itk's lowest mass stars. In Section~\ref{sec:sample} we introduce the criteria that were used to define our sample and follow in Section \ref{sec:data} with a description of the \itk\ data products and the preparation of these data for our following analyses. In Section~\ref{sec:fitting} we outline in detail our treatment of the \itk\ data including a preliminary characterization of the data with outlier rejection and a Markov Chain Monte Carlo parameter estimation. Also in this section we search for transit timing variations (TTVs) in the light curve data that may be due to mutual gravitational interactions within multi-planet systems or other effects, and perform custom fits to the transit shapes of those sources with significantly non-linear ephemerides. We present the full ensemble of transit candidates and stellar parameters in Section~\ref{sec:ensemble}, and conclude in Section~\ref{sec:conclusion}.

\section{Sample of Planet Candidates}
\label{sec:sample}
Our list of cool planet host stars is drawn from a recent \itk\ Object of Interest (KOI) list available through the Exoplanet Archive \citep[][downloaded on September 18, 2014]{Akeson2013}. A total of 4228 planet transit signals toward 3250 targets were selected from the KOI list with dispositions of either ``candidate'' or ``confirmed,'' comprising a high--fidelity catalog of exoplanets \citep[see, {\eg}][]{Morton2011,Fressin2013,Morton2012}. We choose from this list of candidates those with host star color $K_p - J > 2$ and $K_p > 14$ as a cut for M dwarfs \citep{Mann2012}. We also include 6 stars with $r -J > 2.0$ from the study by \cite{Muirhead2014} that pass our red criterion but not our faint criterion: KOI-314, KOI-641, KOI-1725, KOI-3444, KOI-3497, and KOI-4252. Lastly, we also include the new planet discovered by \cite{Muirhead2015}, KOI-2704.03, or \itk-445d.

We cross-match this full list with the list presented by \cite{Muirhead2014} in which near-infrared spectra for 106 stars toward 103 KOIs are presented. Two of the sources in that list are now categorized as false positives: KOI-1459 and KOI-3090. Another binary system, KOI-4463, consists of stars that appear earlier than M0 in \cite{Muirhead2014} and the KOI is not included in the \cite{Dressing2013} catalog. We leave these three targets off our list. We also exclude from further consideration a known M dwarf/white dwarf binary in the list \citep[KOI-256][]{Muirhead2013}, and the giant star KOI-977 \citep{Muirhead2014}. We therefore consider 98 cool KOIs from the \cite{Muirhead2014} list incorporating all 64 targets in the KOI catalog of \citet{Dressing2013}, save one other now-known false positive, KOI-1164. 

The newest release of KOIs postdates both the \cite{Muirhead2014} and \cite{Dressing2013} catalogs, and so we also cross matched our KOI list against the full catalog of \cite{Dressing2013} to find 9 additional cool stars with candidate transit signals: KOI-2480, KOI-2793 KOI-3102, KOI-3094, KOI-4971, KOI-4987, KOI-5228, KOI-5359, and KOI-5692. These targets are some of the smallest and longest period planet candidates in our list and offer exciting possibilities for followup observations.

The final sample we consider for further characterization consists of 167 planet signals toward 107 cool stars observed by \itk. A majority of the stars in this sample (76) show single transit signals, while we find 13 double systems, 10 triple systems, 5 quadruple systems, and 3 quintuple systems. However, the majority of planet candidates, 54.5\%, are in multi-transiting systems. The multi-planet systems have a higher probability of being true planetary systems due to a paucity of astrophysical false positive scenarios that could produce multiple, independent transit-like signals within a single \itk\ aperture \citep[{\eg}][]{Lissauer2014,Rowe2014} while also passing the data validation pipeline \citep{Wu2010}.

\section{Data Preparation}
\label{sec:data}
\subsection{\itk\ Data}
The targets in our sample were observed over the entire course of the \itk\ mission. However, in Quarter 0 only three cool KOI targets were observed. Over the rest of the mission an average of 87\% of the targets in our sample were observed each quarter producing an average of 53,567 long cadence data per target and a total of 5.7 million long cadence photometric measurements for our sample. None of the targets on our sample were observed in short cadence mode until quarter 6 when 9.3\% of the targets made the short cadence target list. This fraction rose fairly steadily for the rest of the mission up to quarter 17 when nearly 25\% of the cool KOIs were observed in short cadence mode producing a total of 25.8 million short cadence data.
\begin{figure*}
\begin{center}
\includegraphics[width=6.5in]{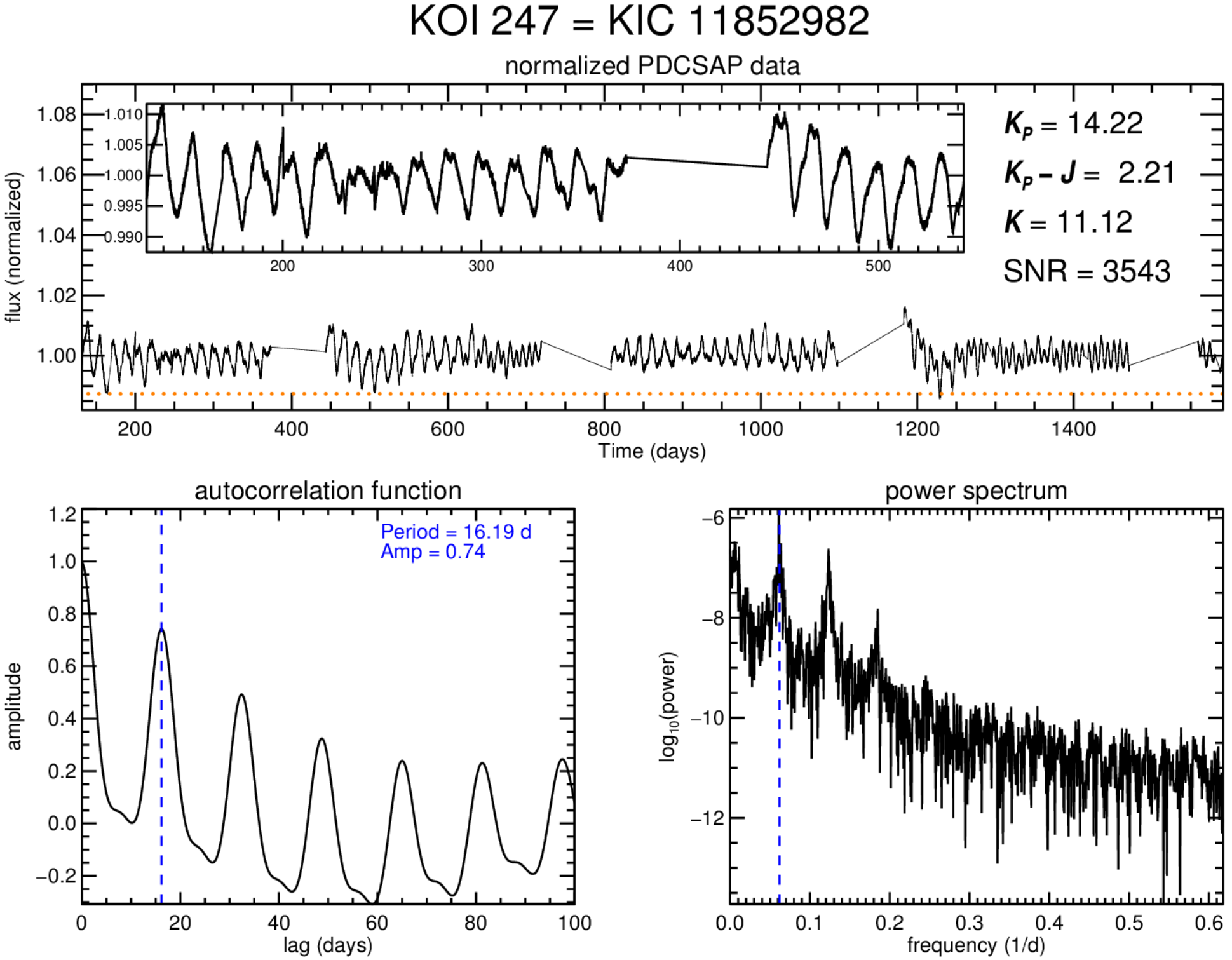}
\caption{\scriptsize{Example of a diagnostic plot for the long cadence data of KOI-247 showing the out of transit data characteristics including the signal to noise of the light curve and absolute photometry. The top panel shows the entire span of the long cadence dataset with a zoom in window of the first 400 days. The transit times are marked on the upper panel plot with colored dots ({\it orange dots} in this example) for reference. The lower panels show periodicities in the out of transit data via the autocorrelation function ({\it lower left}) and Fourier transform ({\it lower right}) from which we estimate the stellar rotation period. The vertical lines ({\it blue, dashed}) denotes the peak of the auto-correlation function and its corresponding frequency.}}
\label{fig:diagnostic}
\end{center}
\end{figure*}

We obtained the light curve data through the the Barbara A. Mikulski Archive for Space Telescopes\footnote{{\url https://archive.stsci.edu/kepler}} (MAST) using Data Release 21 for Quarters 0 through 14, release 20 for Quarter 15, and releases 22 and 23 for Quarters 16 and 17, respectively. For all \itk\ data header keyword definitions, we refer the reader to the \itk\ Archive Manual\footnote{see {\url http://archive.stsci.edu/kepler/manuals/archive\_manual.pdf}}. We consider only data with SAP QUALITY values equal to 0. This excludes data that were taken under non-optimal circumstances or were flagged for other reasons. On average this resulted in a rejection of about 12.5\% of long cadence data per target and 6.2\% of short cadence data per target.

For each KOI, both the Pre-search Data Conditioning \citep[PDCSAP;][]{Stumpe2012,Smith2012} and Simple Aperture Photometry (SAP) data were examined. The SAP data were cotrended using the first 5 cotrending basis vectors available through the MAST website, and then deblended using the FLFRCSAP and CROWDSAP header keywords. In all cases our calibrated SAP data appeared very similar or nearly identical to the PDSCAP data and we default to using the PDCSAP data for all KOIs for the sake of uniformity. 

Before addressing the transit signals, we first look at the raw data for anomalies, trends and other potential problems. Figure~\ref{fig:diagnostic} shows an example of one of our diagnostic plots that displays the entire time series of data, a zoom in of a small portion of the data, and photometry information. A normalized flux series is created for each KOI in our list by concatenating all available data normalized by the median flux value of each quarter. We then subtract the median flux of the combined series and blank out any transit signals using the durations and ephemerides provided by the Exoplanet Archive. These data are then gridded onto a uniform time series and zeroed at values where data were missing. Periods were searched out to 100 days using both an auto-correlation and Fourier transform. The normalized light curves, auto-correlation functions, and spectral power density were then inspected by eye. In a majority of cases where periodic signatures were seen, they are interpreted as modulations due to the combination of stellar rotation and a non-uniform stellar surface brightness.

\subsection{Extracting Transit Signals}
Each of the 167 planet signals described above was extracted from the full \itk\ light curve by fitting a linear drift to the out of transit data extending two transit durations before the beginning of ingress and two durations after egress. The transit times and durations used in this process were taken from the Exoplanet Archive. Linear ephemerides were assumed for each of the transit signals in the extraction process. However, a small buffer of 10\% the reported transit duration was used to account for any potential transit timing variations or errors in the values reported by the Exoplanet Archive. The root-mean-squared (rms) value of the residuals to the linear trend is recorded and applied to all the data from each transit event as the relative flux error. 

\begin{figure*}
\begin{center}
\includegraphics[width=6.5in]{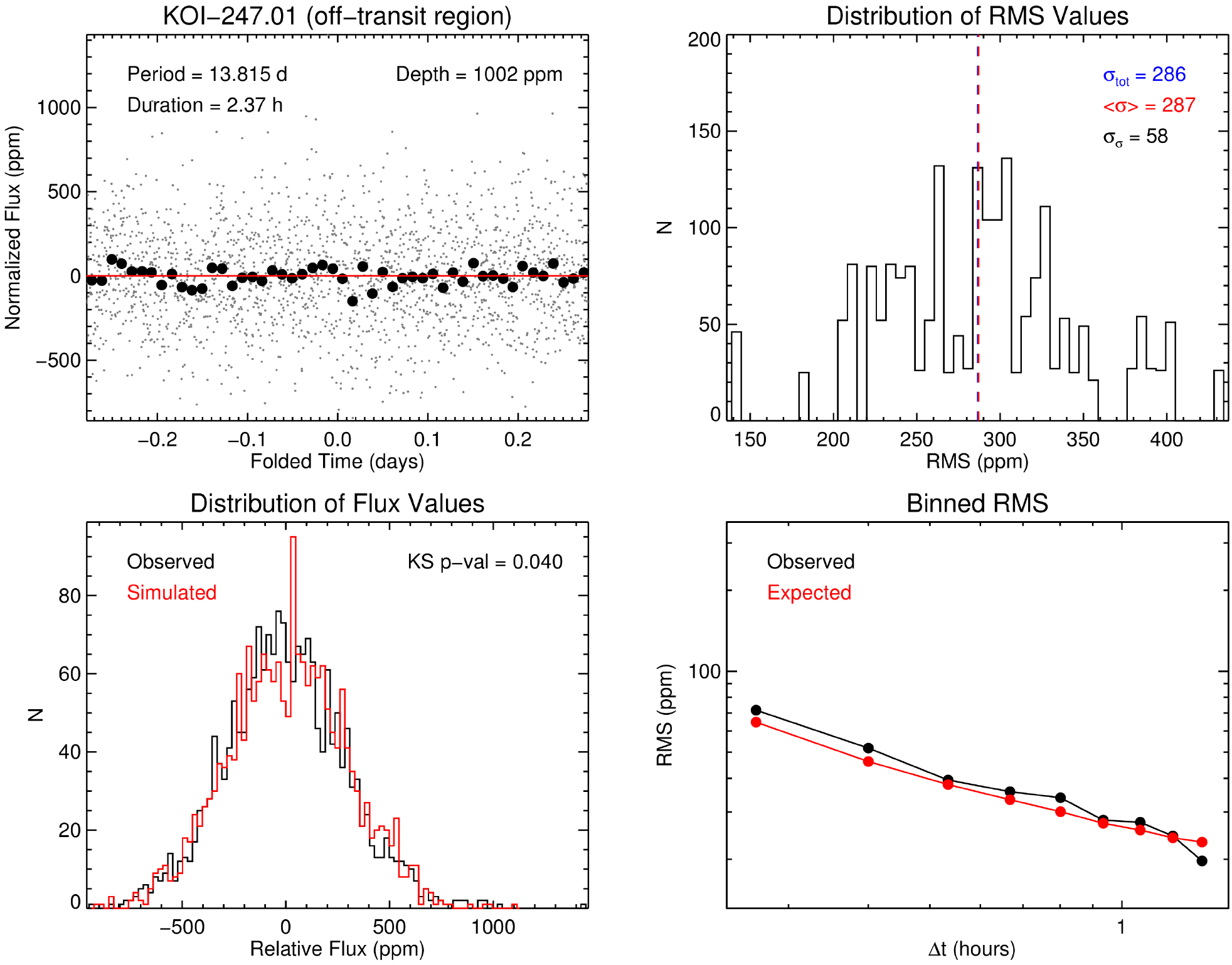}
\caption{\scriptsize{({\it Top left}) The adjacent, transit-free section of the light curve for the specified KOI is shown folded on the period of the planet transit signal. The calibrated \itk\ data are shown as small dots, and binned data are plotted as larger dots to reveal more subtle structure. ({\it Top right}) The distribution of RMS values derived from the detrending process are shown in histrogram form. The RMS of the folded data, $\sigma_{\rm tot}$, is depicted with the blue dotted line; the mean of the RMS values derived from the detrending process, $\langle\sigma\rangle$, is shown as the dotted red line; and the spread in the detrend derived RMS values, $\sigma_\sigma$, is also displayed. ({\it Bottom left}) Histogram of the data from the top left panel is shown and compared with a histogram of values drawn from a normal distribution with zero mean and a standard deviation equal to the RMS of the data. The results from a two-sided Kolmogorov-Smirnov test show the probability that the two distributions were drawn from the same parent sample. ({\it Bottom right}) The phase folded data are binned on a series of time scales, $\Delta t$, starting with the smallest bin that will include at least 20 points and stepping up in 10 bins to one half the transit duration as reported by the \itk\ team. This curve is shown in relation to the expected trend \citep[\eg,][]{Winn2008}}}
\label{fig:noise}
\end{center}
\end{figure*}

Next, each transit signal was confirmed using a box-least squared algorithm \citep[BLS;][]{Kovacs2002} optimized to oversample the projected BLS peak width by a factor of 3 \citep[see][]{Ofir2014}. This typically produced convincing transit signals with durations and ephemerides that were in general agreement with the values on the Exoplanet Archive. However, there were a few exceptions. KOI-1686.01 and KOI-1408.02\footnote{We note that after the time we downloaded the KOI information from the NASA Exoplanet Archive, the disposition for KOI-1408.02 was changed to ``false positive."} do not show a convincing transit signal and we hereafter leave these off of our list. Also, the period reported for KOI-1725 was found to be approximately 9 minutes off, necessitating an independent period search to adequately retrieve this signal. In cases where a transit signal was apparent in long cadence data, but problematic or not clearly seen in the short cadence data (typically due to a paucity of short cadence data), the transit parameters derived from the long cadence data were applied to the short cadence data. Examples of extracted transit signals are shown in Section~\ref{sec:fitting}.

Correlated noise produced by either instrumental or astrophysical phenomena can have a significant affect on the interpretation of astronomical light curve data \citep[see, {\eg}][]{Pont2006,Carter2009}. Therefore, in addition to the transit extraction, a section of the light curve with no transit signal was extracted in exactly the same manner as the transit signal, but according to a mid-transit time advanced by 5 times the reported transit duration. This produced a transit-free section of the light curve immediately adjacent to the extracted transit events. Figure~\ref{fig:noise} shows one example of a ``blank" extraction as well as the basic analyses we use to assess the noise properties of our data (see caption for more details). We find that the distribution of data values for each KOI can be reasonably described by a single parameter, $\sigma$ and compares well with synthetic, Gaussian distributed data (typical KS p-values $\gtrsim 0.01$). The fact that the noise properties of our data sample appear nearly Gaussian can be attributed to a variety of factors. A dominant effect is that the stars in our sample are by design faint, meaning that the photon noise is higher than for the rest of the sample which can mask subtler, correlated phenomena. Also, the astrophysical noise from M dwarf light curves are typically from inhomogeneities in the stellar surface brightness coupled with stellar rotation rather than pulsation modes \citep[see, {\eg}][]{Rodriguez2012}. The stellar rotation timescales are typically much longer than the transit durations, so these effects are adequately corrected with our detrending process. We therefore do not consider the effects of correlated noise in later analyses.

A final step in the preparation of our light curves is outlier rejection. This procedure removes astrophysical ({\eg}, flares) and instrumental effects not accounted for in the above procedures as well as points that were not adequately detrended. We reject outliers from the phase folded transit signal by binning the data into bins that are one half the integration time of the observations or with widths that contain at least 20 data points per bin. From the distribution of data points in each bin a robust estimation of the standard deviation is calculated using the median absolute deviation:
\begin{equation}
{\rm MAD} = {\rm median}\left(|x_i - {\rm median}({\bf x})|\right),
\end{equation}

\noindent where the residuals are given by ${\bf x} = \{x_0, x_1 ... x_n\}$. The MAD is then scaled to estimate the standard deviation assuming a Gaussian distribution so that $\sigma = 1.4826\,{\rm MAD}$ and then data are rejected with absolute deviation from the median beyond a threshold $n\sigma$ where
\begin{equation}
n = \sqrt{2}\,{\rm erf}^{-1}(1-\eta/N)
\end{equation}
where N is the number of data point under consideration. Removing outliers in this manner produces a minimal effect on the statistical properties of the data by removing points that are inconsistent with the original robust estimation of the standard deviation of the sample given the sample size. We use a value of $\eta = 0.1$ which translates to $2.8 \lesssim n \lesssim 4.0$ for our dataset.

\section{Transit Fitting}
\label{sec:fitting}
\subsection{Long and Short Cadence Fits Using a Linear Ephemeris Model}
\label{sub:linearfits}
We characterize our vetted sample of 165 planets around 106 cool stars---now excluding KOI-1686.01 and KOI-1408.02---by first fitting all the long and short cadence data available with a linear ephemeris transit model using a Markov Chain Monte Carlo parameter estimation algorithm. Our light curve model uses the analytic solutions from \cite{Mandel2002} for a quadratic stellar limb darkening law that provides a relative flux model for planet-to-star size ratio, projected separation, and limb darkening parameters. The hyper-geometric functions of those solutions need to be evaluated numerically and present a computational barrier. We therefore use a circular planet orbit to convert time into projected separation for a given period and transit duration. This allows us to side-step solving Kepler's equation, and instead perform the transformation from time to relative separation between the star and planet with simple trigonometric functions. Under this approximation the ingress and egress of the model are exactly symmetric also halving the number of computations needed for each model call. Of course, this disallows subtle effects due to eccentric orbits to be adequately modeled and care must be taken when interpreting the derived transit duration in terms of stellar density \citep{Seager2003,Kipping2010a}. However for our sample, this effect can be accounted for and is expected to have a negligible effect on the derived transit parameters.

For our fits we parametrize our model with the scaled planet radius, $R_p/R_\star$; the impact parameter, $b$; the duration from the first to the fourth contact point of the transit, $\tau_{\rm tot}$; the time of mid-transit as measured nearest to the middle of the \itk\ light curve, $t_0$; the period, $P$; and two limb-darkening parameters, $q_1$ and $q_2$, that characterize the full range of quadratic parameter space of monotonically decreasing and positive value profiles \citep{Kipping2013e}. 



Before our models can be compared to data, the effect of finite integration times must be considered \citep[{\eg}][]{Kipping2010b,Price2014}. The \itk\ Mission produced time series data sampled at two different intervals using a single exposure time. The exposure time (accumulated time of flux from a celestial source on a given pixel) is $t_{\rm exp} = 6.020$\,s, and for every exposure there is a fixed CCD readout time of $t_{\rm read} = 0.519$\,s. The short cadence data is made up of 9 such exposures and therefore the time between the start of successive short cadence data is $(t_{\rm exp} + t_{\rm read})\times9 = 58.849$\,s. However, the time interval over which the astronomical signal is integrated is one read shorter than this, {\ie}, $t_{\rm smooth}^{\rm short} = 9t_{\rm exp} + 8t_{\rm read} = 58.330$\,s. Similarly, the long cadence data are made up of 270 integrations and therefore the time between successive integration times is $t_{\rm cadence}^{\rm long} = 1765.463$\,s and the smoothing time $t_{\rm smooth}^{\rm long} = 1764.944$\,s. 

To account for the effects of integration time, we first calculate the planet path across the stellar disk assuming the planet is in a circular orbit using $b = a\,\cos(i)/R_\star$. The light curve for this planetary trajectory is oversampled and then smoothed using a uniform filter of width $t_{\rm smooth}$. This is analogous the resampling procedure recommended by \cite{Kipping2010b}, and we hereafter refer to this process as resampling. The degree of resampling needed to produce an accurate model using this method will depend on the transit parameters. Therefore we numerically determine the optimal resampling for each transit candidate based on the parameters from preliminary fits enforcing an resampling of at least 5. For a grid of transit parameter values spanning the full range of $R_p/R_\star$ and $\tau_{\rm tot}$ in our data set, and for an impact parameter of 0 (the effect of finite integration time is most severe for low impact parameter transits), we first calculate a reference transit model resampled by a factor of 3001. We then calculate transit curves for the same set of input parameters resampled in steps of 2 from 3 to 501. The smallest resampling value that produces peak to peak discrepancies with the reference model of less than one part per million is then recorded. We then construct a grid of values from this procedure that we use to interpolate the optimal resampling values to be used for any of our targets based on their preliminary transit parameters. 
\begin{figure}
\begin{center}
\includegraphics[width=0.99\columnwidth]{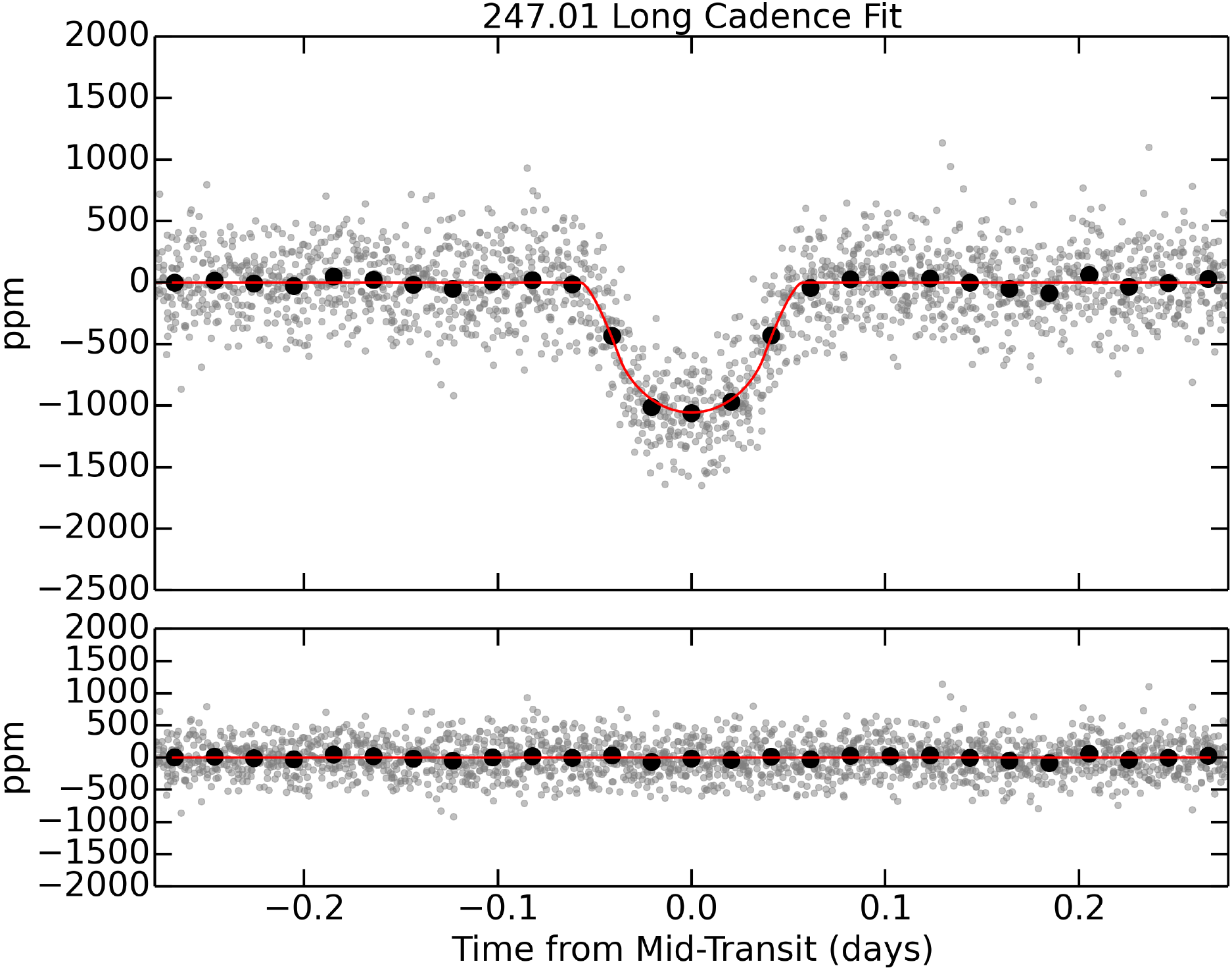}
\caption{\scriptsize{Phase folded long cadence data for KOI-247.01 are shown as gray dots. These data binned at a timescale approximately equal to the original sampling of the long cadence data stream are shown as black dots for viewing purposes only. The best fit model is shown in red and the residuals of this fit are shown in the bottom panel. }}
\label{fig:longtransit}
\end{center}
\end{figure}
\begin{figure}
\begin{center}
\includegraphics[width=0.99\columnwidth]{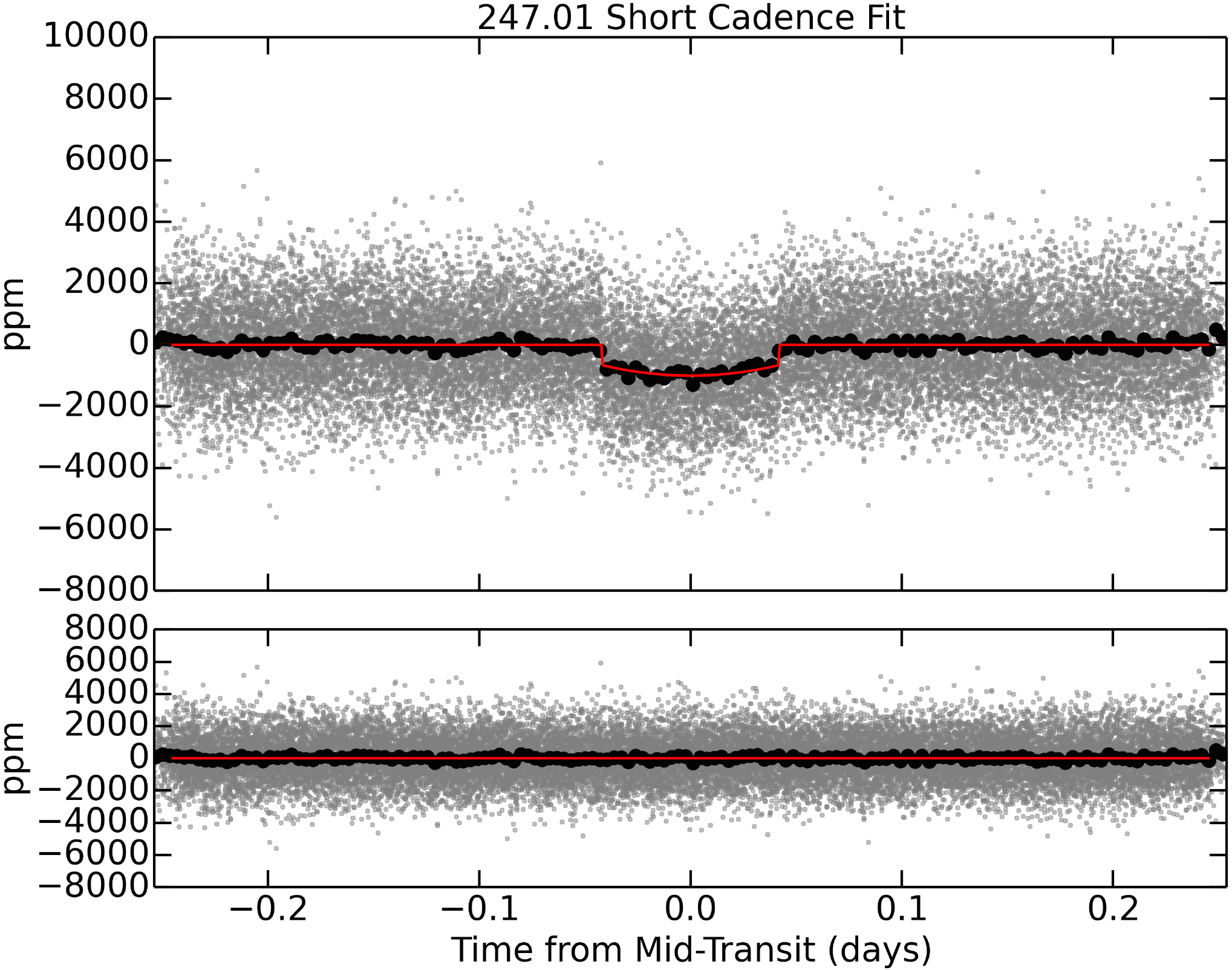}
\caption{\scriptsize{Same as Figure~\ref{fig:longtransit}, but for the KOI-247.01 short cadence data.}}
\label{fig:shorttransit}
\end{center}
\end{figure}

\begin{figure*}
\begin{center}
\includegraphics[width=6.5in]{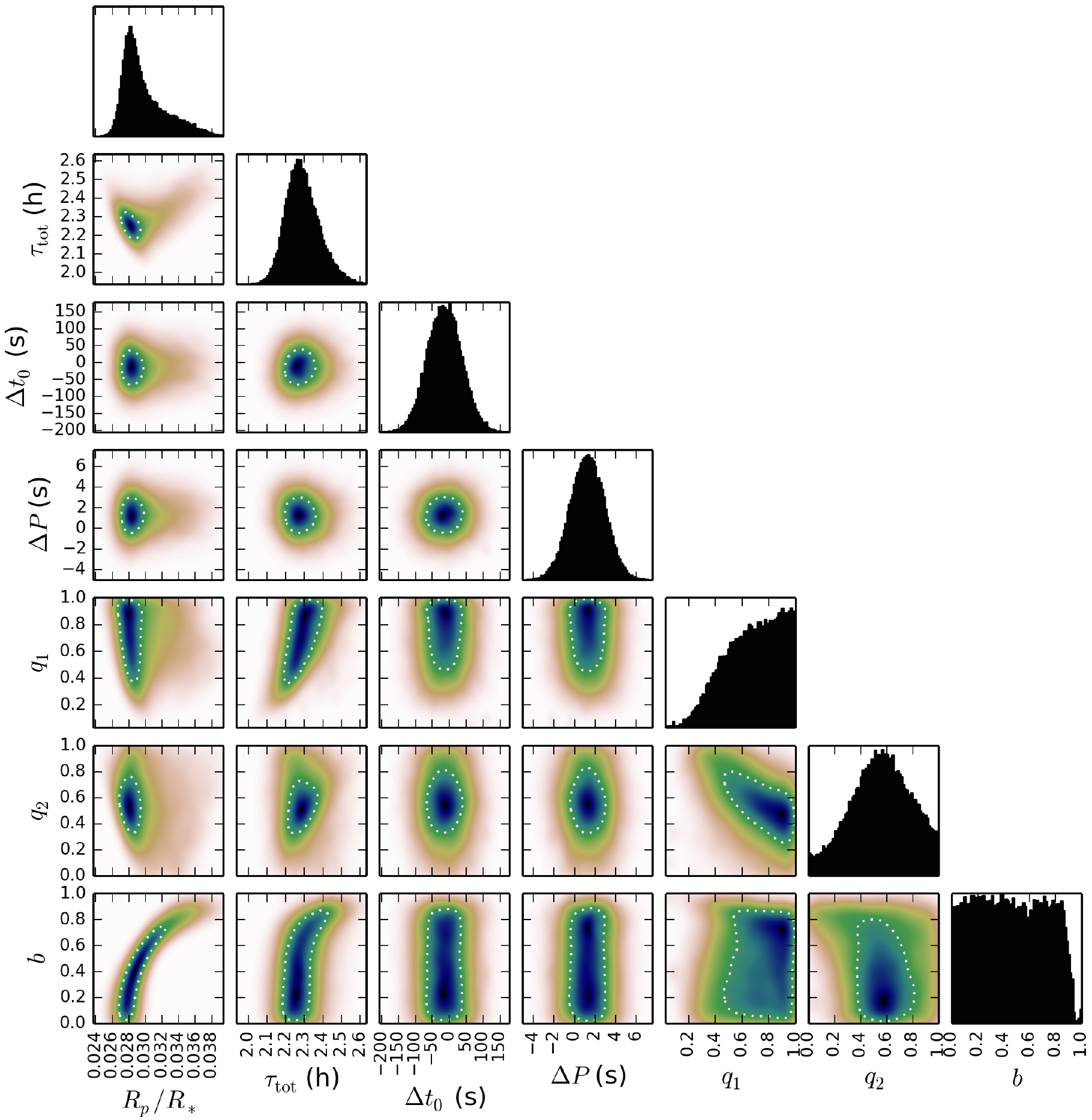}
\caption{\scriptsize{Array of 1-d and 2-d posteriors for the long cadence fit shown in Figure~\ref{fig:longtransit}. The 2-d posteriors were constructed using a 2-d kernel density estimation that reveals covariances between parameters, most notably for $R_p/R_\star$, $b$, and $\tau_{\rm tot}$.}}
\label{fig:triangle}
\end{center}
\end{figure*}

We use a Bayesian framework to determine the best fit values for our 7 model parameters and their associated errors. To evaluate the likelihood, we do not resample the model at each data timestamp. Instead we phase fold the data at each trial period, $P$, and mid transit time, $t_0$, and interpolate our resampled model to the phase folded timestamps of the data. This speeds up each likelihood call by an order of magnitude or more. The quantity $(R_p/R_\star)^2$ is a scale parameter in the problem and we therefore apply a Jefferys prior to this parameter. We note that this has a small to negligible affect on our posterior samples as we are data-dominated rather than prior-dominated for the majority of our transit candidates. Each of the other free parameters have uniform priors ({\ie}, no prior). 

We use the {\tt emcee} affine invariant Markov Chain Monte Carlo ensemble sampler \citep{Foreman-Mackey2013} with 1000 chains, or ``walkers'' ($n_w = 1000$). The initial values of each walker were over-dispersed in most parameters based on the estimated values found by fitting the transit shape with a quick and flexible Levenberg-Marquardt fitting algorithm \citep{Markwardt2009}. The relative planet radius, \rprs, is dispersed in a uniform manner from 0 to a factor of 2 larger than value obtained from the preliminary fit; the full duration, $\tau_{\rm tot}$, is dispersed from half to twice the preliminary fit value; the impact parameter, $b$, is dispersed uniformly from 0 to 1; the period, $P$, is dispersed by $\pm 1$ second from the nominal value; the mid transit times, $t_0$, span 2 minutes uniformly; and the limb darkening parameters, $q_1$ and $q_2$ are dispersed uniformly between 0 and 1.

The walkers are evolved for $n_b = 1000$ steps and then analyzed. We use the correlation length, $c_l$, to assess if the chains have reached a sufficiently mixed state. The burn-in stage was re-run with a larger number of steps if the number of independent draws, $n_bn_w/c_l$, was found to be less than 10,000. The sampler was then reset and the walkers restarted from their last location for an additional 1000 steps. These last 1000 steps for each 1000 walkers ($10^6$ samples total) comprise the final posterior samples that we use to estimate the transit parameters.

The results of the long cadence data fits are summarized in Table~\ref{tab:lcfit}, and an example fit can be seen in Figure~\ref{fig:longtransit}. The median values for planet period, mid transit time, relative radius, duration and impact parameter are reported along with the half width of the shortest $1\,\sigma$ interval of the posterior for each parameter. These values are a useful reference. However, they conceal details about the probability of the these parameter values. Figure~\ref{fig:triangle} shows a series of the 2-dimensional posterior probability distributions for the 7 free parameters in the fits. The expected covariance between parameters such as the impact parameter, $b$, and the relative size of the planet, $R_p/R_\star$, can be clearly seen. The MCMC chains are available for download such that these parameter dependencies can be properly accounted for in future statistical studies.

For 79 transit signals toward 36 cool KOIs there exist short cadence data. We follow the exact procedure outlined above for these data including preliminary fits and MCMC posterior sampling. These results are summarized in Table~\ref{tab:scfit}. The short cadence data fit for the same KOI shown in Figure~\ref{fig:longtransit}, KOI-247.01, is shown in Figure~\ref{fig:shorttransit} for reference.

\subsection{Transit Timing Variations}
\label{sec:TTV}
\subsubsection{TTV Search}
\label{subsub:TTVSearch}
\begin{figure}
\begin{center}
\includegraphics[width=0.99\columnwidth]{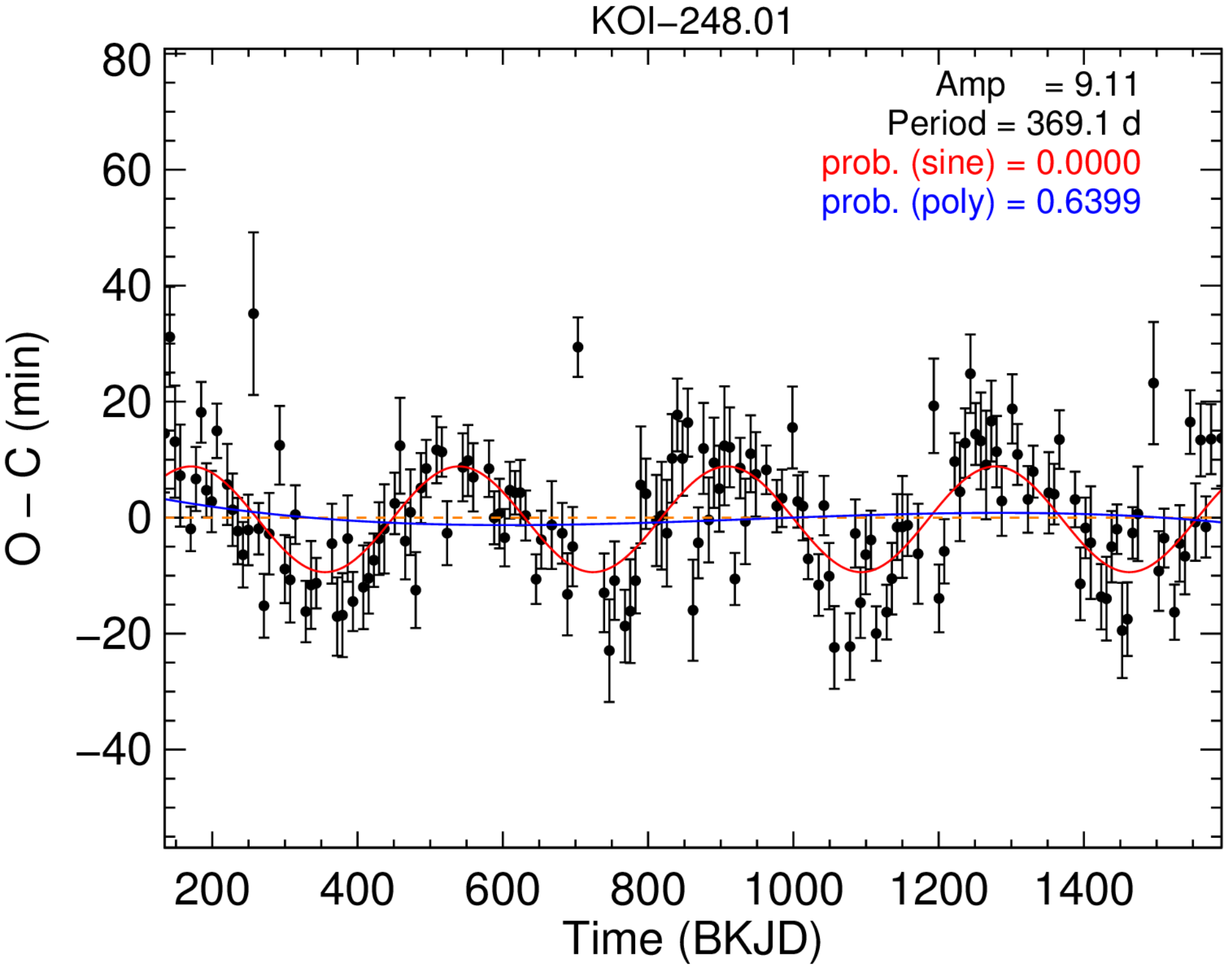}
\caption{\scriptsize{The transit timing variations ($O-C$) of KOI-248.01 fit with a pure sinusoid ({\it red}) and a polynomial ({\it blue}). These fits are only used for assessing the significance of a potential TTV signal and are not used in the fitting of the transits (see Section~\ref{subsub:TTVFit}).}}
\label{fig:TTVs}
\end{center}
\end{figure}

For each transit signal we use the best-fitting transit model to fit for the times of each transit event in search of potential transit timing variations (TTVs). A single parameter, $\Delta t_0$ quantifies the time deviation of mid-transit in relation to the expected time based on a linear ephemeris from the best fits. The model light curve is fit to each transit event letting only $\Delta t_0$ float using a Levenberg-Marquardt minimization \citep{Markwardt2009} to produce a list of observed-minus-calculated ($O-C$) values corresponding to each transit. Figure~\ref{fig:TTVs} shows an example of one of the known TTV planets in our sample, KOI-248.01.

To assess the significance of potential TTV signals we first calculate the RMS scatter in the times of mid-transit as estimated by the median absolute deviation $\sigma_{O-C}$ and compare that to the median value of the estimated errors on the transit times $\bar{\sigma}_{TT}$ \citep{Mazeh2013}. We consider values of $\sigma_{O-C}/\bar{\sigma}_{TT} > 3.0$ as significant. Next we compute a Lomb Normalized Periodogram\footnote{\url{http://www.exelisvis.com/docs/LNP_TEST.html}} for the calculated $O-C$ transit times. We calculate a $p$-value for this peak by producing 10,000 periodograms for the $O-C$ data randomly scrambled. The fractional number of periodogram peaks in the simulation that are greater than or equal to the original peak is interpreted as the probability that the measured periodogram is due to random noise, $p_{\rm LNP}$. This probability value is considered significant when $p_{\rm LNP} \leq 0.001$.

Lastly, we fit both a sine curve and a polynomial to the $O-C$ data. The sine curve model contains an amplitude, period, phase and offset. The starting parameters for the fit are a 1 minute amplitude, a period equal to the location of the peak of the periodogram, and zero phase and offset. To assess the significance of the fit results for the polynomial and sine curve models, we perform an F-test on the fitted parameters by comparing the $\chi^2$ values and degrees of freedom from a single parameter fit (a mean) and the polynomial or sine model. We again consider $p_{\rm sine} \leq 0.001$ and $p_{\rm poly} \leq 0.001$ significant.

\subsubsection{TTV Results}
The results were scrutinized by eye to weed out TTV signals due to stroboscopic effects and other, non-dynamical processes \citep{Szabo2013}. The results from our TTV search are summarized in Table~\ref{tab:ttv}. We recover 12 KOIs with significant TTV signals, 11 of which are in multi-transiting systems. These 12 planet candidates comprise 7.2\% of the full M dwarf planet candidate sample and they are found toward 7 of the 106, or 6.6\% of all M dwarf KOIs. All our TTV detections have been detected previously and are reported in the literature \citep{Mazeh2013,Wu2013,Kipping2014a}. However, these new transit timing results use all data from the \itk\ mission. Following are a few notes regarding our TTV search.

KOI-3284 is reported to have a significant TTV signal by \cite{Kipping2014a}. Our tests show a signal at a period of $\sim 190$\,d in both the periodogram and the sinusoid fit. However, the false alarm probability of the periodogram peak is found to be very high and this KOI also failed our F-test for the sinusoidal fit. Therefore we do not include this planet candidate in our list. KOI-2306 has $\sigma_{O-C}/\bar{\sigma_{TT}} = 3.12$ due to the under sampling of the transit by the long cadence data and we therefore exclude it. KOI-1907 and KOI-2130 show some signs of long period TTV signals at $\sim 700$\,d and $\sim 1100$\,d periods, respectively. But both these signals fall narrowly below our selection criteria and are therefore excluded.

KOI-952.02 is not reported by \cite{Mazeh2013} as a significant TTV source. However, we find that in 17 quarters of data the periodicity at $\sim 260$ days is significant. This matches the period reported by \cite{Fabrycky2012b}. KOI-952.01 does not produce a signal significant enough to warrant inclusion in our list, though we do find that the first 8 quarters of data are consistent with the results of \cite{Fabrycky2012b}, and the period of $\sim 260$\,d is apparent in our periodogram as the second highest peak but with a high formal false positive probability.

\subsubsection{Fitting Transit Signals with TTVs}
\label{subsub:TTVFit}
Transit timing variations can significantly affect the perceived transit shape under the assumption of a linear ephemeris. The effect is to essentially smear out the ingress and egress and potentially fill in the depth of the transit. The details depend on the exact nature of the TTVs. However it is typical that TTVs will bias the impact parameter to higher values, the transit duration to larger values, and the limb darkening parameters will tend toward values that produce a more severe contrast between the center of the star and the limb. 

Due to these effects, we refit the transit signals in our sample that show significant TTVs after folding on the individual transit times derived above. We first reject any individual transits that have mid transit time errors that are either ill defined or are larger than 2\,$\sigma$ from the median error. We then perform a transit fit with the same model outlined in Section~\ref{sec:fitting} except that instead of fitting the period and mid-transit time, we fix the individual transit times. 

\begin{figure}
\begin{center}
\includegraphics[width=0.99\columnwidth]{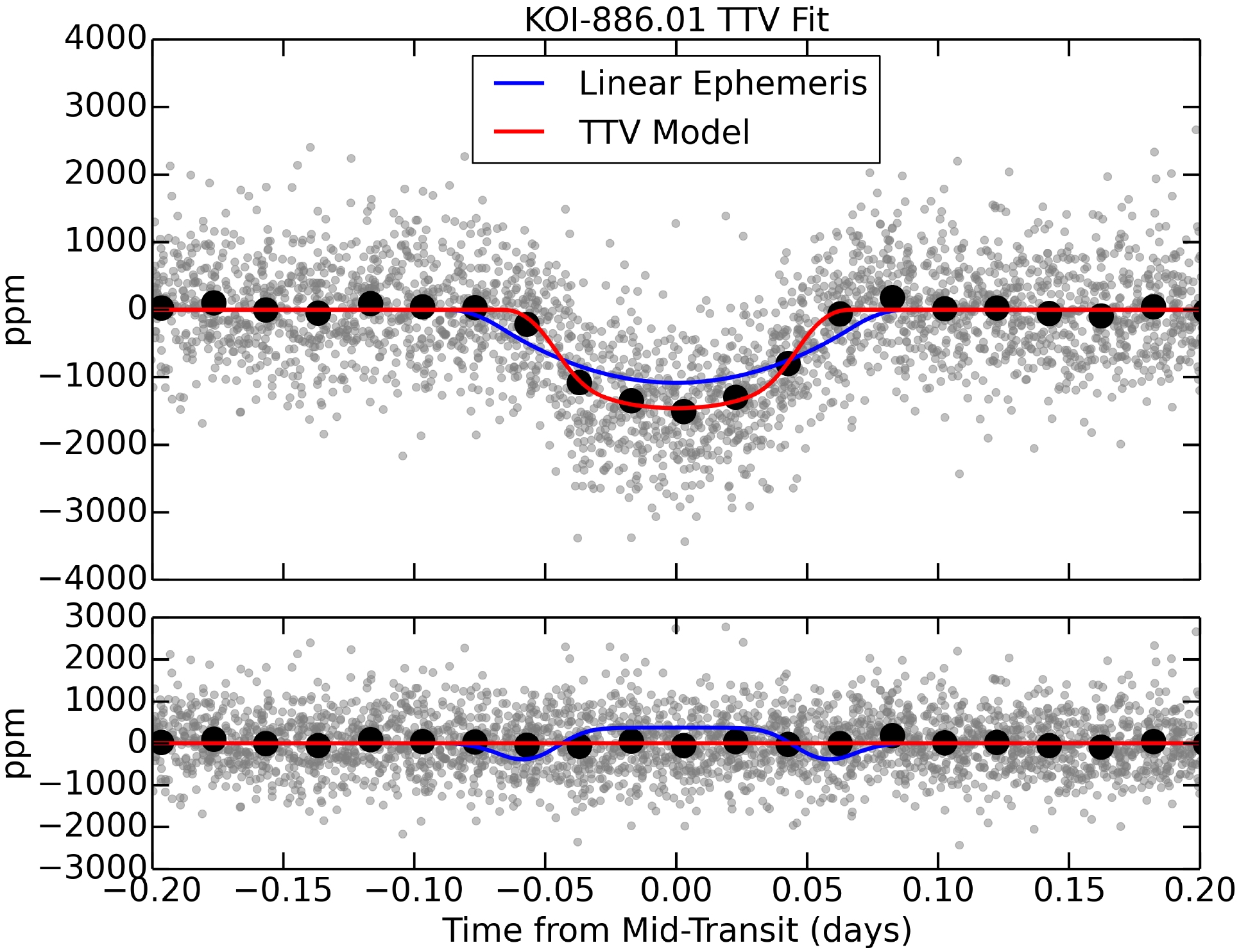}
\caption{\scriptsize{({\it top}) Long cadence \itk\ photometry of KOI-886.01 phase folded on the transit times derived in Section~\ref{sec:TTV}. The best fit model assuming a linear ephemeris is shown in blue, the best fit model for the data folded on the non-linear transit times is shown in red. ({\it bottom}) The residuals of the best fit non-linear model. The difference between the linear and non-linear model is shown in blue. Assuming a linear ephemeris for this target that shows peak-to-peak TTVs of $\sim 2$ hours significantly affect the derived transit parameters, in particular the transit duration.}}
\label{fig:TTVcomp}
\end{center}
\end{figure}

We choose a large TTV source, KOI-886.01, as an example showing the potential effects of fitting linear transit model to a planet that displays significant TTVs. The $\sim 2$\,hr peak-to-peak TTVs for KOI-886.01 bias the fits toward a larger impact parameter, a smaller planet, and a longer duration. The median posterior values for the impact parameter and relative planet size are discrepant at the 0.3, and 1.2\,$\sigma$ levels. However the derived transit durations are in disagreement with 98\% confidence. These results are shown pictorially in Figure~\ref{fig:TTVcomp}.

\section{False Positive Probability}
\label{sec:FPP}
The \itk\ pipeline is known to have produced a high fidelity sample of transiting exoplanets \citep{Wu2010,Morton2011,Morton2012,Christiansen2013,Fressin2013}. Up to this point we have treated every signal as a transiting exoplanet. However, it is prudent to assign to each transit signal a probability that the signal was generated from another astrophysical scenario. We use the methods of \cite{Morton2011} and \cite{Morton2012} to analyze the light curves shapes that we have extracted to assign a false positive probability (FPP) of each transit signal independently.

\begin{figure*}
\begin{center}
\includegraphics[width=6.5in]{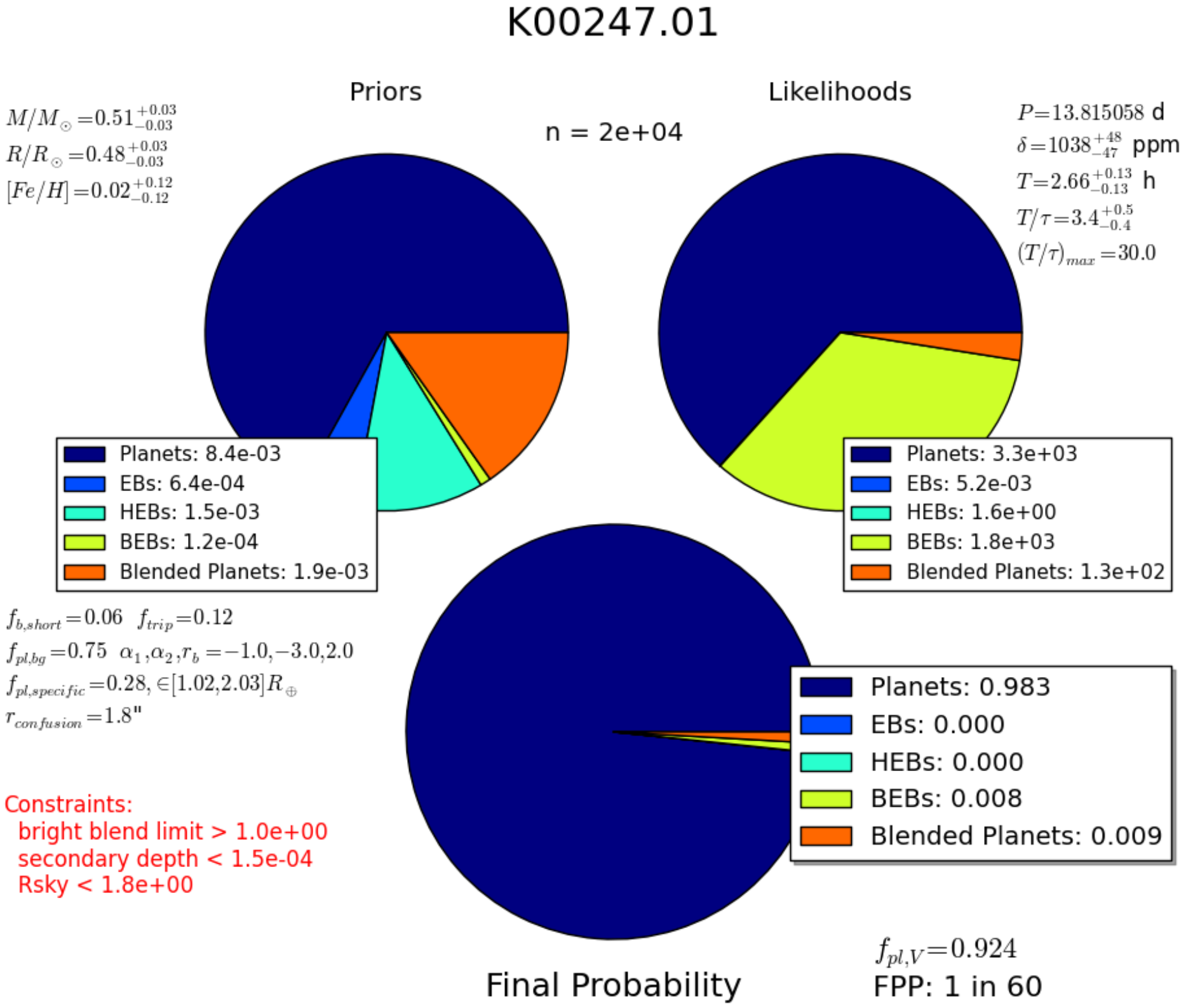}
\caption{\scriptsize{Diagnostic plot showing a the key results of the false positive probability analysis for the sample of transit planet cadidates around low mass stars. The top left pie chart shows the priors likelihoods of the five different scenarios considered: Transiting planet ({\it Planets}), eclipsing binary ({\it EB}), heirarchical eclipsing binary ({\it HEB}), background eclipsing binary ({\it BEB}), and blended planet. These fractions are calculated with a Galactic model in the direction of the target star with an assumed planet occurrence ($f_{pl,specific}$). The top right is the likelihood of these different scenarios given the shape of the long cadence light curve. For this case, KOI-247, the signal is most likely a transit signal around the intended star. However, the most likely false positive scenarios are background eclipsing binaries and blended planet signals.}}
\label{fig:FPP}
\end{center}
\end{figure*}

These FPPs are reported in Table~\ref{tab:FPP} along with the probability of the transiting planet scenario compared to all other astrophysical scenarios, $P = L_{\rm TP}/L_{\rm FP}$; the specific occurrence assumed in the calculation,$f_{pl,specific}$; and the specific planet occurrence needed to achieve a threshold FPP of 0.005, $f_{p,V}$. Included in each calculation is also a confusion radius within which false positives are permitted to exist.  For this radius we use three times the uncertainty in the multi-quarter difference-image pixel response function (PRF) fit that is reported at the Exoplanet Archive [the ``PRF $\Delta\theta_{\rm MQ}$ (OOT)'' column].  The minimum exclusion radius we allow is 0.5 arcseconds, and the default value we use if no value is available is 4 arcseconds.   An example of a diagnostic plot generated by the FPP analysis is shown in Figure~\ref{fig:FPP}.  

We find that 11\% of the sample, or 18 of the 165, has a FPP of larger than 10\%, consistent with estimations of the entire \itk\ sample \citep{Morton2011,Fressin2013}. However, 6 of these high FPP targets are either known planets in the literature (\eg, KOI-254.01 \citep{Johnson2011}, and KOI-886.02 \citep{Steffen2013}), KOI-1422.05 \citep{Rowe2014}) or are part of 3 or 4 transit systems much less likely to be a false positives. Therefore this is a high fidelity sample of transiting exoplanets around the lowest mass stars observed by the \itk\ primary mission.

We do note that our treatment of exclusion radius ignores the possibility of more distant PRF contamination, as detected via the period-epoch match study of \cite{Coughlin2014}, which found that ``parent'' eclipsing stars even up to 10-100 arcseconds from the target star were able to cause ``child'' false positive signals.  While that work discovered over 600 false-positive KOIs, it also highlighted the possibility of further distant contaminants that might remain undetected because the ``parent'' may not be a known eclipsing system.  

In order to estimate the rough probability of any of the present KOIs being false positives via this mechanism, we may use the numbers discussed by \cite{Coughlin2014}.  That work identified 12\% of all known KOIs (not all planet candidates) to be from PRF contamination.  However, they pointed out that only about 1/3 of the stars in the \itk\ field were downloaded, so it might be reasonable to assume that for every discovered PRF contaminant, there might be two undiscovered, putting the overall rate at about 36\%.  According to this reasoning, about 24\% of all KOIs might be PRF contaminants unable to be discovered by the period-epoch match method.  

However, they also go on to point out that 5/6 of the FPs they detected were also identified as FPs by other methods (e.g., pixel-centroid offets, detected secondary eclipses, etc.).  So this implies that of those previously mentioned 24\%, only 1/6 of those, or 4\% of all KOIs, might be long-distance PRF contaminants undetected by any \itk\ FP vetting procedure and thus making it to planet candidate status.  Comparing this to the $\sim$64\% of all KOIs expected to be true planets, we estimate that an additional $\sim$6-7\% of \itk\ planet candidates, beyond what we calculate here using the methods of \citet{Morton2012}, could still be false positives.  Incorporating in detail this additional long-distance PRF contamination into quantitative models of false-positive probability is thus warranted, but beyond the scope of this present work.

Additionally, we also note that the FPPs presented in this paper do not consider the number of independent transit signals in the light curve, nor the possibility of detected TTVs, both of which may substantially reduce the FPP \citep[{\eg},][]{Lissauer2014,Rowe2014,Ford2011}.  

\section{The ensemble of M dwarf planet candidates}
\label{sec:ensemble}
The cool KOI catalog enables study of the smallest and possibly most numerous planet population discovered by \itk\ and helps to advance our knowledge of planet formation around the most common types of stars. It is estimated that 75\% of the stars within 10\,pc are M dwarfs \citep{Henry1994,Reid2002,Henry2004}. Therefore by targeting this population we are also learning what can be expected of the closest planetary systems outside our Solar System.  

\begin{figure*}
\begin{center}
\includegraphics[width=6.5in]{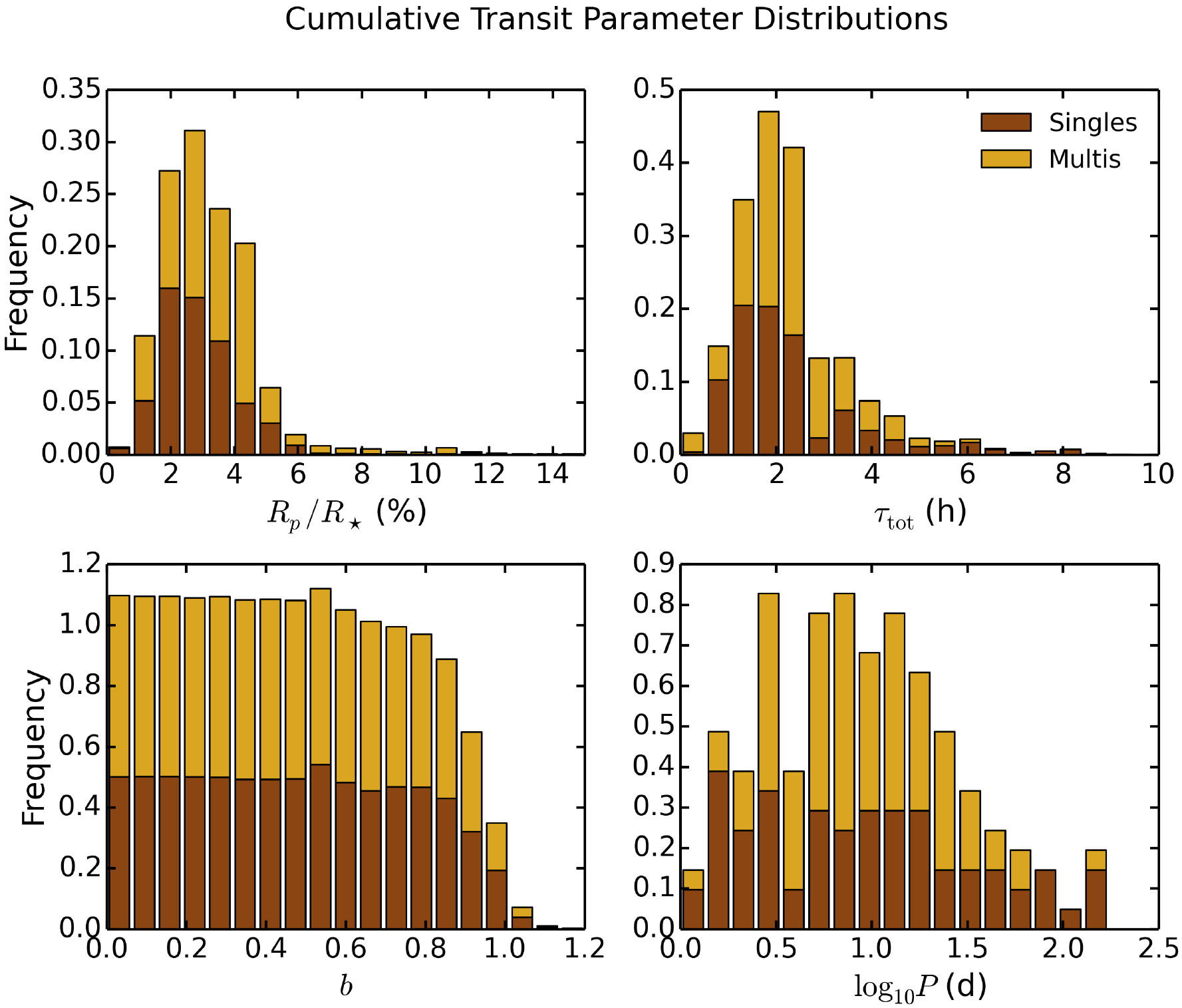}
\caption{\scriptsize{Cumulative distributions of four of the seven transit parameters for the sample of exoplanet candidates orbiting \itk's coolest dwarf stars. The radii of the planet candidates ({\it top left}) are displayed in terms of a percentage of the radius of the host star. The total duration (first to fourth contact point, {\it top right}) is shown in units of hours. The impact parameter ({\it bottom left}) is seen to be mostly indeterminable from the long cadence data except for KOI-254/\itk-45 which accounts for the bump near $b=0.54$. The periods of the planet candidates span more than 2 orders of magnitude and are shown on a $\log_{10}$ scale ({\it bottom right}) to reveal more detail of the distribution. The stacked histograms differentiate the sample of single transit systems ({\it brown}) and planets in multi-transit systems ({\it gold}).}}
\label{fig:cum}
\end{center}
\end{figure*}

To further our understanding of this sample of small planets we present uniformly derived transit parameters for all known transit signals around the cool KOIs. These stars constitute a small fraction (about 2\%) of the total \itk\ targets. However, the sample is large enough to allow for meaningful statistical analyses \citep{Morton2014a,Ballard2014}. Since M dwarf stars are difficult to characterize observationally, it is also important that our sample is small enough such that each individual star can be addressed with followup observations. 
 
The planet candidates of this work have been drawn from the Exoplanet Archive list using the cool dwarf photometric cuts of \cite{Mann2012}. Additional vetting was performed using near infrared, medium resolution spectroscopy \citep{Muirhead2012a, Muirhead2014}. Our final sample contains 165 planets around 106 cool stars. The number of single transit systems totals 77, while there are 11 double systems, 10 triple systems, 5 quadruple systems, and 3 quintuple systems. A total of 53.3\% of these planets are found in multi-transit systems, and of these multis 11.4\% show significant TTV signals. On the contrary, only one single transit system out of 77, or 1.3\%, shows a significant TTV signal. 

The final results of our transit fits to \itk\ long and short cadence data are summarized in Tables~\ref{tab:lcfit} and \ref{tab:scfit}, respectively. These tables display the results from the linear ephemeris model for all KOIs except for those listed in Table~\ref{tab:ttv}. For those sources we report the period, $P$, and mid transit time, $t_0$, from the linear ephemeris fits (although it should be noted that these parameters are not strictly defined in this context) and the other transit parameters from the non-linear ephemeris fits. An earlier version of this catalog has already been used in the literature to infer statistical properties of the \itk\ M dwarf planet population \citep{Morton2014a}, and is presented here such that it may be used for further statistical studies. Each transit signal has been treated individually, and we have generated posterior samples of the 7 transit parameters using uninformed priors that are available for download along with a suite of diagnostic plots for each KOI.

The cumulative distributions for the four transit parameters that have the most direct relevance to the planet statistics are displayed in Figure~\ref{fig:cum}. For this plot and those that follow we distinguish between the planets that are in single transit and multi-transit systems.
\begin{figure*}
\begin{center}
\includegraphics[width=6.5in]{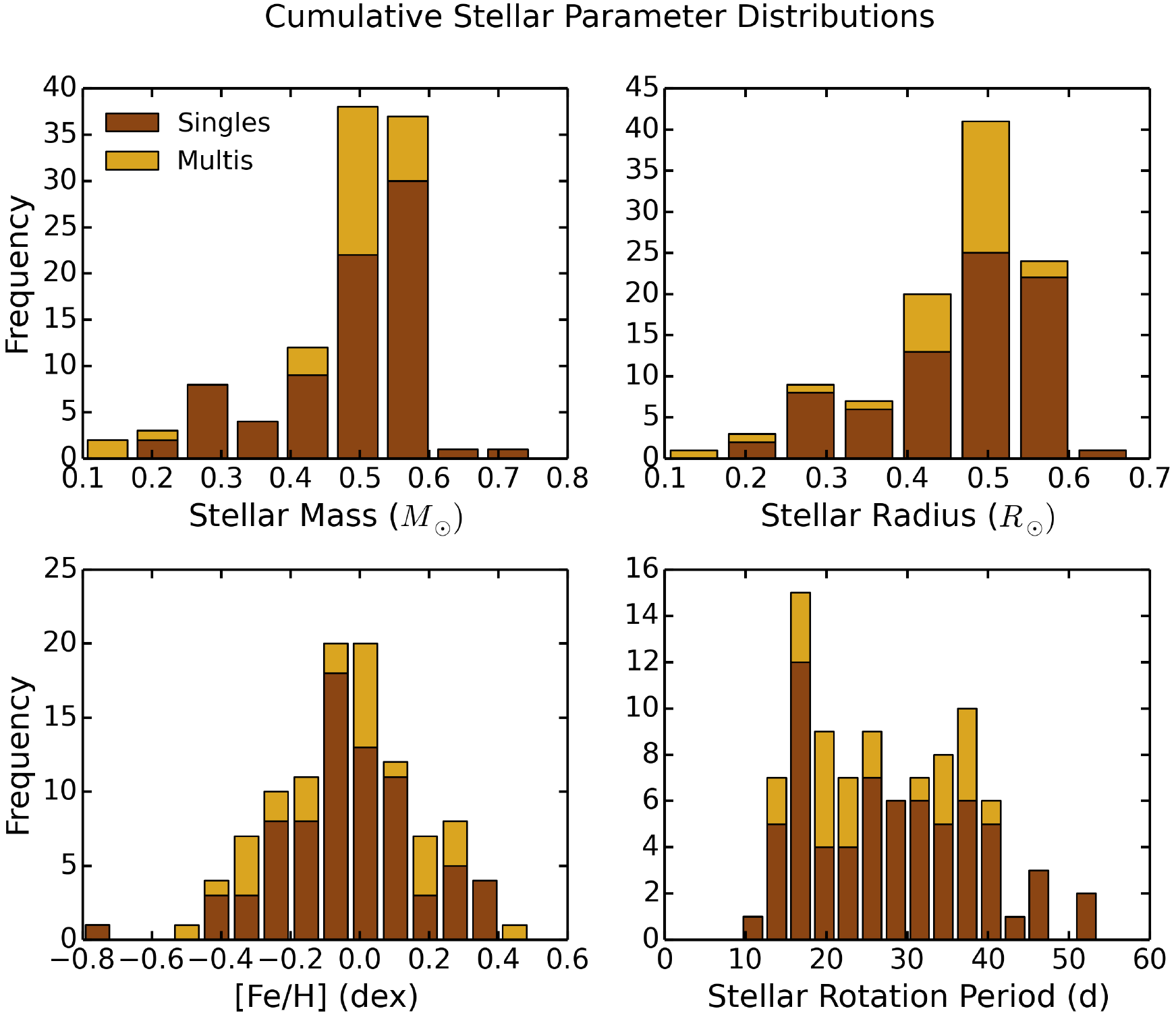}
\caption{\scriptsize{Distribution of stellar parameters for the final ensemble of 106 cool KOIs. The host stars of single transit systems and multi-transit systems have been distinguished are shown in dark and light shading, respectively.}}
\label{fig:star}
\end{center}
\end{figure*}

\subsection{Stellar Characteristics}
\label{sec:stars}
The physical parameters of the transiting planets are intimately tied to the stellar parameters. We therefore also consolidate data for the stellar sample both from this work and from the literature. Stellar masses, radii and effective temperatures were obtained from the lists of \cite{Muirhead2014} and \cite{Dressing2013}. By default we use the stellar parameters derived from the medium resolution, infrared spectroscopy of \cite{Muirhead2014}. The method uses a calibrated empirical relationship between the shape of the pseudo-continuum in the $K$-band spectrum  to infer a stellar effective temperature \citep[H$_2$O--K2 index][]{Rojas2012}. The equivalent widths of the Ca\,{\sc i} triplet and Na\,{\sc i} doublet within the same band are used to estimate the stellar metallicity using a relationship calibrated on nearby wide binaries with FGK type stars \citep{Rojas2010}. The mass and radius of the star is then estimated by interpolating these \teff\ and \mh\ values onto stellar evolutionary tracks \citep{Dotter2008,Feiden2011}. 

For KOIs that do not have parameters derived with near infrared spectra, we use the stellar parameters from \cite{Dressing2013}. Here the authors interpolate the wide band photometry from the \itk\ input catalog \citep[KIC;][]{Brown2011} onto stellar evolution models to obtain masses, radii and metallicities. The mass and radius values derived by this method are typically in reasonable agreement with \cite{Muirhead2014}, while the metallicity estimates are comparatively less reliable.

These compiled values and errors are presented in Table~\ref{tab:star} along with the photometry from the KIC. In addition to this information we also include our estimate of the stellar rotation period derived from the rotational modulation of an inhomogeneous surface brightness distribution. We are able to detect this rotational signature in a large fraction of our sample, about 86\%, and report the period corresponding to the largest peak of the auto-correlation function that we validate by visual inspection. The stellar rotation period can be an important parameter in the characterization of the planet sample as this allows for age estimates \citep{Barnes2003} as well as activity levels \citep[{\eg},][]{Reiners2012}. The distribution of stellar parameters are shown for host stars of single and multi-transit systems in Figure~\ref{fig:star}.

\section{Summary and Conclusion}
\label{sec:conclusion}
Many exciting discoveries and insights from the \itk\ Mission have come from the relatively small sample of M dwarf stars \citep{Muirhead2012a, Muirhead2013, Johnson2011, Johnson2012, Dressing2013, Morton2014a,Ballard2014, Quintana2014}. The small sizes of these stars make it easier to probe deeper into the realm of super-Earth and terrestrial planets where planets form most readily. The cool surface temperatures facilitate detections of ever smaller planets in or near to where liquid water may exist on their surfaces due to the shorter orbital periods and higher transit probability. While this sample is a mere 2\% of the total number of stars \itk\ observed during its primary mission, it offers a glimpse into the formation of the most numerous planets orbiting the most numerous stars in the Galaxy. 

These facts have played a large role in motivating our group's efforts to understand this population of stars and planets. In this work we present a uniform analysis of the photometry of cool dwarf stars spanning the full \itk\ primary mission the results of which are catalogs of transit parameters and stellar parameters for 165 transit candidates orbiting 106 low-mass dwarf stars. The stellar parameters are taken primarily from \cite{Muirhead2014}, and supplemented with values from \cite{Dressing2013}. We add new stellar rotation periods estimated directly from the \itk\ light curves, and recover rotational modulation for nearly 86\% of our targets. 

As the statistical treatments of the \itk\ data set continue to advance and improve, these transit parameters are meant to serve as a valuable dataset. To facilitate further studies we make available the posterior distributions of the transit parameters for each planet candidate including short cadence fit parameters where available. Diagnostic plots for each KOI created during the reduction and analysis of the light curves are also available for each star and transit. 

\acknowledgements
JJS would like to thank Jason Eastman, David Kipping, Ellen Price, and Natalie Batalha for their helpful input regarding various aspects of this work. All of the data presented in this paper were obtained from the Mikulski Archive for Space Telescopes (MAST). STScI is operated by the Association of Universities for Research in Astronomy, Inc., under NASA contract NAS5-26555. Support for MAST for non-HST data is provided by the NASA Office of Space Science via grant NNX13AC07G and by other grants and contracts. This paper includes data collected by the \itk\ mission. Funding for the \itk\ mission is provided by the NASA Science Mission directorate. A.V. and B.T.M are supported by the National Science Foundation Graduate Research Fellowship, Grants No. DGE 1144152 and DGE 1144469, respectively.

\clearpage
\LongTables
\begin{deluxetable}{ccccccccccc}
\tablewidth{0pt}
\tabletypesize{\scriptsize}
\tablecaption{\label{tab:lcfit} Transit Parameters for Long Cadence Fits}
\tablehead{\colhead{KOI} &
\colhead{$P$} &
\colhead{$\delta P$} &
\colhead{$t_0$} &
\colhead{$\delta t_0$} &
\colhead{$R_p/R_\star$} &
\colhead{$\delta R_p/R_\star$} &
\colhead{$\tau_{\rm tot}$} &
\colhead{$\delta \tau_{\rm tot}$} &
\colhead{$b$} &
\colhead{$\delta b$} \\
\colhead{} &
\colhead{days} &
\colhead{sec} &
\colhead{BJD-2454833} &
\colhead{sec} &
\colhead{\%} &
\colhead{\%} &
\colhead{hrs} &
\colhead{min} &
\colhead{} &
\colhead{} }
\startdata
                         247.01 &  13.815050 &     1.67 &   858.062058 &    50.70 &    2.909 &    0.223 &    2.2848 &     5.328 &  0.45 &  0.31 \\
248.01\tablenotemark{$\dagger$} &   7.203854 &     0.51 &   861.856433 &    30.48 &    4.105 &    0.111 &    2.5224 &     2.808 &  0.36 &  0.24 \\
248.02\tablenotemark{$\dagger$} &  10.912760 &     1.56 &   868.268258 &    57.14 &    3.338 &    0.234 &    2.5104 &     5.544 &  0.38 &  0.28 \\
                         248.03 &   2.576571 &     0.22 &   860.071548 &    36.42 &    2.677 &    0.207 &    1.7088 &     4.104 &  0.47 &  0.32 \\
                         248.04 &  18.596146 &     4.61 &   866.516301 &    93.95 &    2.766 &    0.183 &    2.2488 &     6.336 &  0.46 &  0.32 \\
                         249.01 &   9.549275 &     0.43 &   863.305081 &    17.86 &    4.046 &    0.205 &    1.5960 &     2.880 &  0.45 &  0.30 \\
                         250.01 &  12.283005 &     0.79 &   858.246284 &    27.39 &    5.290 &    0.394 &    2.9016 &     7.560 &  0.71 &  0.23 \\
250.02\tablenotemark{$\dagger$} &  17.251179 &     1.73 &   839.934988 &    43.47 &    4.367 &    0.266 &    1.9608 &     5.184 &  0.51 &  0.29 \\
\enddata
\tablenotetext{$\dagger$}{Transit parameters derived from fits to individual transit times. Period and mid-transit time values are taken from fits assuming a linear ephemeris.}
\end{deluxetable}

\clearpage
\LongTables
\begin{deluxetable}{ccccccccccc}
\tablewidth{0pt}
\tabletypesize{\scriptsize}
\tablecaption{\label{tab:scfit} Transit Parameters for Short Cadence Fits}
\tablehead{\colhead{KOI} &
\colhead{$P$} &
\colhead{$\delta P$} &
\colhead{$t_0$} &
\colhead{$\delta t_0$} &
\colhead{$R_p/R_\star$} &
\colhead{$\delta R_p/R_\star$} &
\colhead{$\tau_{\rm tot}$} &
\colhead{$\delta \tau_{\rm tot}$} &
\colhead{$b$} &
\colhead{$\delta b$} \\
\colhead{} &
\colhead{days} &
\colhead{sec} &
\colhead{BJD-2454833} &
\colhead{sec} &
\colhead{\%} &
\colhead{\%} &
\colhead{hrs} &
\colhead{min} &
\colhead{} &
\colhead{} }
\startdata
                         247.01 &  13.815048 &     2.76 &   816.617339 &    34.72 &    2.932 &    0.090 &    2.0760 &     1.944 &  0.36 &  0.24 \\
248.01\tablenotemark{$\dagger$} &   7.203857 &     0.68 &  1063.565399 &    27.28 &    4.511 &    0.128 &    2.7024 &     2.448 &  0.82 &  0.03 \\
248.02\tablenotemark{$\dagger$} &  10.912775 &     2.26 &  1064.699385 &    66.70 &    3.142 &    0.099 &    2.8056 &     2.016 &  0.24 &  0.18 \\
                         248.03 &   2.576568 &     0.18 &  1063.620783 &    23.69 &    2.710 &    0.133 &    1.6512 &     1.800 &  0.48 &  0.26 \\
                         248.04 &  18.596038 &     5.24 &  1071.074246 &    96.71 &    2.610 &    0.191 &    2.3616 &     4.824 &  0.46 &  0.31 \\
                         249.01 &   9.549278 &     0.74 &   853.755337 &    13.69 &    3.950 &    0.097 &    1.5936 &     1.008 &  0.29 &  0.20 \\
250.01\tablenotemark{$\dagger$} &  12.282930 &     1.07 &  1067.057752 &    24.86 &    5.526 &    0.186 &    2.9400 &     2.664 &  0.81 &  0.05 \\
250.02\tablenotemark{$\dagger$} &  17.251180 &     2.63 &  1046.949800 &    45.36 &    4.824 &    0.434 &    2.1336 &     4.752 &  0.76 &  0.12 \\
\enddata
\tablenotetext{$\dagger$}{Transit parameters derived from fits to individual transit times. Period and mid-transit time values are taken from fits assuming a linear ephemeris.}
\end{deluxetable}

\clearpage
\begin{deluxetable}{ccccccccc}
\tablewidth{0pt}
\tabletypesize{\scriptsize}
\tablecaption{\label{tab:ttv} M dwarf Planets with Transit Timing Variations}
\tablehead{\colhead{KOI} &
\colhead{$N$} &
\colhead{$\sigma_{O-C}/\bar{\sigma}_{\rm TT}$} &
\colhead{LNP amp.} &
\colhead{$p_{\rm LNP}$} &
\colhead{sine amp.} &
\colhead{$P_{\rm TTV}$} &
\colhead{$p_{\rm sine}$} &
\colhead{$p_{\rm poly}$} \\
\colhead{} &
\colhead{} &
\colhead{} &
\colhead{} &
\colhead{} &
\colhead{(min)} &
\colhead{(days)} &
\colhead{} &
\colhead{} }
\startdata
                         248.01 &  159 & {\bf 2.19} & 29.15 & {\bf 0.0001} &   9.71 &  365.97 & {\bf 0.00000} &       0.68916  \\
                         248.02 &  100 & {\bf 2.08} &  7.25 & {\bf 0.0009} &  15.06 &  365.76 &       0.00383 &       0.42931  \\
                         250.01 &   95 &       1.86 & 22.62 & {\bf 0.0001} &   9.76 &  743.74 & {\bf 0.00000} &       0.03606  \\
                         250.02 &   52 &       1.63 &  9.38 & {\bf 0.0009} &   7.60 &  809.02 & {\bf 0.00007} &       0.18420  \\
                         314.01 &   71 &       1.85 & 12.29 & {\bf 0.0001} &   5.46 & 1111.01 & {\bf 0.00000} & {\bf 0.00000}  \\
                         314.02 &   50 &       1.88 & 14.91 & {\bf 0.0001} &  13.86 & 1022.36 & {\bf 0.00000} & {\bf 0.00000}  \\
                         314.03 &  117 &       1.56 &  5.93 &       0.2357 &  32.43 & 1402.62 & {\bf 0.00008} & {\bf 0.00031}  \\
                         463.01 &   59 &       1.32 &  9.28 &       0.0021 &   3.65 &  314.51 & {\bf 0.00025} &       0.94999  \\
\enddata
\end{deluxetable}

\clearpage
\begin{deluxetable}{ccccc}
\tablewidth{0pt}
\tabletypesize{\scriptsize}
\tablecaption{\label{tab:FPP}False Positive Probability Results }
\tablehead{\colhead{KOI} &
\colhead{FPP} &
\colhead{$P$} &
\colhead{$f_{p,specific}$} &
\colhead{$f_{p,V}$}}
\startdata
                                                   247.01 &  0.0165 &    215 & 0.276 &    0.92400 \\
248.01\tablenotemark{$\dagger$}\tablenotemark{$\ddagger$} &  0.0000 & 290321 & 0.218 &    0.00069 \\
248.02\tablenotemark{$\dagger$}\tablenotemark{$\ddagger$} &  0.0902 &     39 & 0.258 &    5.08000 \\
                         248.03\tablenotemark{$\ddagger$} &  0.0004 &   8541 & 0.276 &    0.02330 \\
                         248.04\tablenotemark{$\ddagger$} &  0.0018 &   1998 & 0.276 &    0.09960 \\
                                                   249.01 &  0.0000 & 848734 & 0.269 &    0.00023 \\
250.01\tablenotemark{$\dagger$}\tablenotemark{$\ddagger$} &  0.0940 &     64 & 0.149 &    3.08000 \\
250.02\tablenotemark{$\dagger$}\tablenotemark{$\ddagger$} &  0.0069 &    707 & 0.202 &    0.28200 \\
\enddata
\tablenotetext{$\dagger$}{Source of significant TTV signal.}
\tablenotetext{$\ddagger$}{Multi-transit candidate system.}
\end{deluxetable}

\clearpage
\begin{deluxetable}{cccccccccccc}
\tablewidth{0pt}
\tabletypesize{\scriptsize}
\tablecaption{\label{tab:star} M Star {\em Kepler} Objects of Interest}
\tablehead{\colhead{KOI} &
\colhead{KIC} &
\colhead{$N_{\rm pl}$}  &
\colhead{Mass (\msun)} &
\colhead{Radius (\rsun)} &
\colhead{$T_{\rm eff}$ ($K$)} &
\colhead{[Fe/H] (dex)} &
\colhead{Ref.} &
\colhead{$P_{\rm rot}$ ($d$)} &
\colhead{$K_p$} &
\colhead{$J$} &
\colhead{$K_{\rm s}$}}
\startdata
 247 &   11852982 & 1 & $0.51^{+0.03}_{-0.03}$ & $0.48^{+0.03}_{-0.03}$ & $3735^{+ 49}_{- 33}$ & $ 0.02^{+0.12}_{-0.12}$ & 1 &  16.2 & 14.22 & 12.01 & 11.12 \\
 248 &    5364071 & 4 & $0.55^{+0.04}_{-0.04}$ & $0.52^{+0.04}_{-0.04}$ & $3838^{+111}_{- 74}$ & $-0.02^{+0.14}_{-0.14}$ & 1 &  18.3 & 15.26 & 13.18 & 12.38 \\
 249 &    9390653 & 1 & $0.40^{+0.05}_{-0.05}$ & $0.39^{+0.04}_{-0.04}$ & $3562^{+ 50}_{- 64}$ & $-0.13^{+0.13}_{-0.13}$ & 1 &  43.6 & 14.49 & 12.00 & 11.15 \\
 250 &    9757613 & 4 & $0.55^{+0.05}_{-0.05}$ & $0.52^{+0.05}_{-0.05}$ & $3884^{+199}_{- 27}$ & $-0.13^{+0.13}_{-0.13}$ & 1 &  17.8 & 15.47 & 13.41 & 12.63 \\
 251 &   10489206 & 2 & $0.53^{+0.03}_{-0.03}$ & $0.50^{+0.03}_{-0.03}$ & $3811^{+ 48}_{- 71}$ & $-0.06^{+0.11}_{-0.11}$ & 1 &  14.5 & 14.75 & 12.48 & 11.68 \\
 252 &   11187837 & 1 & $0.52^{+0.03}_{-0.03}$ & $0.49^{+0.03}_{-0.03}$ & $3745^{+ 70}_{- 56}$ & $ 0.06^{+0.11}_{-0.11}$ & 1 &  39.5 & 15.61 & 13.42 & 12.55 \\
 253 &   11752906 & 2 & $0.59^{+0.04}_{-0.04}$ & $0.56^{+0.04}_{-0.04}$ & $3759^{+182}_{- 34}$ & $ 0.49^{+0.14}_{-0.14}$ & 1 & \nodata & 15.25 & 13.09 & 12.29 \\
 254 &    5794240 & 1 & $0.58^{+0.03}_{-0.03}$ & $0.55^{+0.03}_{-0.03}$ & $3793^{+133}_{- 34}$ & $ 0.32^{+0.13}_{-0.13}$ & 1 &  15.8 & 15.98 & 13.75 & 12.89 \\
\enddata
\end{deluxetable}

\end{document}